\documentclass{article}
\usepackage{spconf,amsmath,graphicx}

\usepackage{enumitem}
\setlist{nosep, leftmargin=14pt}

\usepackage{mwe} 

\title{Enabling Collagen Quantification on HE-stained Slides Through Stain Deconvolution and Restained HE-HES.}
\name{%
\begin{tabular}{@{}c@{}}
Guillaume Balezo$^{1}$ \qquad 
Christof A. Bertram$^{2}$ \qquad 
Cyprien Tilmant$^{3}$\\ 
Stéphanie Petit$^{4}$ \qquad 
 Saima Ben Hadj$^{1}$ \qquad 
Rutger H.J. Fick$^{1}$
\end{tabular}}
\address{$^{1}$Tribun Health, Paris, France, $^{2}$University of Veterinary Medicine Vienna, Austria,\\ $^{3}$GHICL, Lille, France, $^{4}$Xpath Nord, Leulinghem, France.}
\begin{document}
\maketitle

\begin{abstract}
In histology, the presence of collagen in the extra-cellular matrix has both diagnostic and prognostic value for cancer malignancy, and can be highlighted by adding Saffron (S) to a routine Hematoxylin and Eosin (HE) staining. However, Saffron is not usually added because of the additional cost and because pathologists are accustomed to HE, with the exception of France-based laboratories. In this paper, we show that it is possible to quantify the collagen content from the HE image alone and to digitally create an HES image. To do so, we trained a UNet to predict the Saffron densities from HE images. We created a dataset of registered, restained HE-HES slides and we extracted the Saffron concentrations as ground truth using stain deconvolution on the HES images. Our model reached a Mean Absolute Error of 0.0668 $\pm$ 0.0002 (Saffron values between 0 and 1) on a 3-fold testing set. We hope our approach can aid in improving the clinical workflow while reducing reagent costs for laboratories.

\end{abstract}
\section{Introduction}
\label{sec:intro}

HE is the primary tissue stain worldwide in histology and is the foundation for pathological diagnosis of many diseases. The Hematoxylin stains cell nuclei a purplish blue and Eosin stains the extracellular matrix and cytoplasm pink, allowing visualization of different cellular and extracellular structures. In French laboratories, Saffron is commonly added to the HE stain to highlight collagen fibers in the connective tissue of the extracellular matrix with an orange hue (Figures 1 and 3 show examples of HE and HES images). Visualization of collagen is beneficial for diagnosis of many tumor and non-tumor disease processes. In oncology, this stain allows better visualization contrasting the connective tissue stroma against the tumor cells and enable quantification of the stroma which has been shown to have prognostic value \cite{stroma_cancer}. In some liver diseases, the quantification of the fibrosis - which is mostly collagen - also have a prognostic value \cite{saf_score}. It is therefore of great interest to develop a deep learning model which accurately estimates the collagen from an HE slide, facilitating diagnosis at lower cost.

As it can be seen in Figure 3, the visual cues for collagen are present in the HE slides, but they are not easily identifiable as these features share the stain color of the Eosin. It has been shown that it is possible to reliably predict special stains from HE using conditional Generative Adversarial Networks (cGANs) on registered restained slides \cite{gan_stain}. Thus, we expect it is also feasible to accurately estimate the collagen from HE, because the Saffron only highlights the patterns that are already there. However, in contrast to \cite{gan_stain}, we are interested not only in visualizing a generated HES image as virtual stain, but also to quantify the spatial collagen concentration. Our aim is to estimate the pixel-wise collagen content through Saffron that can be extracted from the images via stain deconvolution, instead of the RGB pixel values. Finally, to display the predictions in a way the pathologists are accustomed to, we can add the estimated Saffron concentrations to the HE images and reconstruct an HES image.

\section{Materials and Methods}
\label{sec:majhead}

In this section, we present our methodology for creating a deep learning model that can predict pixel-wise collagen concentration from an HE image and use it to virtually create an HES image. First, in Section 2.1. we describe our dataset. Then, in section 2.2 our HE-HES registration approach. In 2.3 our stain deconvolution approach to create the ground truth collagen concentration map from the HES, and in 2.4 our deep learning approach. Finally, in 2.5 we explain how we virtually stain the HES.

\subsection{Dataset Acquisition}
\label{ssec:data}

Our dataset \footnote{The dataset was created at the University of Veterinary Medicine Vienna with ethical permission.} is composed of seven canine surgical samples of the breast cancer. First, routine HE slides were created and the slides were scanned using a 3DHistech Panoramic Scan II with an objective magnification of 20x (scan resolution: 0.5$\mu m/$pixel). Then the cover slid of the HE slides was removed and the slides were incubated in Saffron solution as per standard protocol and the HES slides were scanned. Finally, we get a dataset with seven pairs of slides with a total tissue surface of 4.5 $\text{cm}^2$.

\subsection{Slide Registration}
\label{ssec:registration}

The re-staining process results in an image that has minimal tissue deformations and that mostly can be registered by using only rigid transformations (rotations and translations) between pairs of HE-HES slides. To get the correct Saffron concentrations associated with each pixel of the HE slides, we performed robust HE-HES WSI registration using an open-source Quad-Tree-based method \cite{qtree}, and created pixel-perfect matching HE-HES patches using a secondary affine registration applied to each patch location.

\subsection{Ground Truth extraction via stain deconvolution}
\label{ssec:staindeconv}

As the Saffron stain intensity in HES slides is proportional to the collagen content, we consider the prediction of Saffron stain intensity as a proxy task for collagen content. However, in the HES tiles the information of the different stains is mixed up in the 3 color channels and we need to extract the ground truths saffron concentrations $H_{S} \in [0;1]^{n}$ from these multi-stain (H-E-S) images. To do so, we used stain deconvolution \cite{color-deconv} which relies on the empirical Beer Lambert Law that states that there exists a linear relationship in the optical density space between the staining concentrations and the RGB intensities. Let $I_{HES} \in [\![ 0~;~255 ]\!]^{n \times 3}$ be the flattened HES RGB image and $H_{HES} \in R_{+}^{n \times r}$ the corresponding staining densities, with $n$ being the number of pixels and $r$ the number of stains (for HES $r=3$) and $I_0 \in [\![ 0~;~255 ]\!]$ the illuminating light intensity. We can write the following equations:

\begin{equation}
I_{HES}=I_0 \exp (-W H_{HES})\text{, with }  W \in R^{3 \times r}
\end{equation}
\begin{equation}
V_{HES}=W H_{HES} \text{, with } V_{HES}=-\log (\frac{I_{HES}}{I_0})
\end{equation}

$H_{HES}$ can be easily calculated using the Moore-Penrose pseudo inverse of the stain matrix $W$. We used the Non-negative Matrix Factorization (SNMF) method defined in \cite{vahadane} for estimating jointly $W$ and $H_{HES}$. We estimate a unique stain matrix for each slide, instead of different matrices for each patch, by using a random set of tiles in the tissue allowing to learn a robust global stain matrix and to ensure consistency of the Saffron extraction over the WSIs. Once $H_{HES}$ estimated, we extracted the Saffron concentrations by keeping the corresponding channel. Inspired by \cite{vahadane}, we normalized the intensities by dividing them by the pseudo maximum Saffron concentration at 99\% computed over the WSI.

\subsection{Predicting Saffron concentration from HE}
\label{ssec:unet}

We trained a UNet \cite{unet} architecture with a Mobilenet-v2 \cite{mobilenetv2} backbone pretrained on Imagenet to predict the Saffron concentrations $H_{\hat{S}} \in [0;1]^{n}$. The optimization process was done with a final ReLU activation using a Weighted Mean Squared Error (wMSE Loss). The ReLU activation was used because the Saffron concentrations belong to the optical density space which is a linear space with only positive values. We used the wMSE loss to compensate for the unbalanced distribution of pixels in the Saffron stained areas ($H_{S} > 0$) versus the rest of the image: we computed the proportions of the two classes on the training set and for each pixel the associated loss was weighted by the proportion of the opposite class. The optimization was done during 50 epochs using the Adam optimizer with a constant learning rate of 1e-4 and a stopping criterion on the validation loss. We used a batch size of 16 with 512x512 pixels patches at 20x. To avoid over-fitting, we also used a weight decay of 1e-5 and data augmentation with geometrical transformations (flipping, rotations, shear) and stain color augmentation \cite{stain_augment}. We used a three folds validation on the seven slides resulting in an average of 8.5k testing patches. For each fold, the training was done using the training slides with 80\% of the patches (17k patches) as training data and the other 20\% as validation data (1.5k patches).

\subsection{HE to HES Reconstruction}
\label{ssec:he2hes}

To provide the pathologists a realistic-looking HES image $I_{HE+\hat{S}}$ based on quantitative collagen content without directly predicting the RGB pixels, we defined a reconstruction method based on stain deconvolution. Let $\hat{H}_{S}$  be the Saffron concentration predicted by our model, $I_{HE}$ be the corresponding input HE image and $H_{HE}$ its HE concentration. We assume that we estimated $W_{HE}$ on the HE slide and $W_{S}$ on an arbitrary HES slide. If we note $[., .]$ the concatenation operator, we have:

\begin{equation}
H_{\tilde{E}} = where (H_{\hat{S}} > H_{E} + \epsilon, H_{E} = 0)
\end{equation}

\begin{equation}
V_{HE+\hat{S}} = [W_{HE}, W_{S}]H_{H\tilde{E}\hat{S}}
\end{equation}

We can get $I_{HE+\hat{S}}$ from $V_{HE+\hat{S}}$ using (2). In (3), we model the fact that on HE images Saffron areas have strong Eosin concentrations whereas on HES images the paired regions have strong Saffron concentration only. Therefore, for pixels with high Saffron density estimated, we set to zero the Eosin concentration, otherwise the resulting RGB color would be close to red (mix between Eosin and Saffron) instead of the desired orange hue. The margin parameter $\epsilon$ was manually set to 0.1. Note that this method was empirically designed and can likely be improved upon by simulating the chemical staining process more precisely. One of the advantages of this method, compared to generative models predicting the HES RGB intensities, is that with stain deconvolution we can reconstruct the RGB image while only modifying the color distribution of the image where the Saffron is predicted.

\section{Results}
\label{sec:results}

In this section, we present the different results of our method. First, in Section 3.1. we report the different metrics of our deep learning model on the pixel-wise regression task and we show the strong capability of the model to quantitatively estimate the collagen. Then in 3.2, we qualitatively analyze the performances of the model by looking at predictions and the reconstructed HES images in some different areas of the tissue. Figures 1, 2 and 3 are linked by the test ROIs that are drawn in Figure 1.

\begin{figure}[htb]
\begin{minipage}[b]{.48\linewidth}
  \centering
  \centerline{\includegraphics[width=3.5cm]{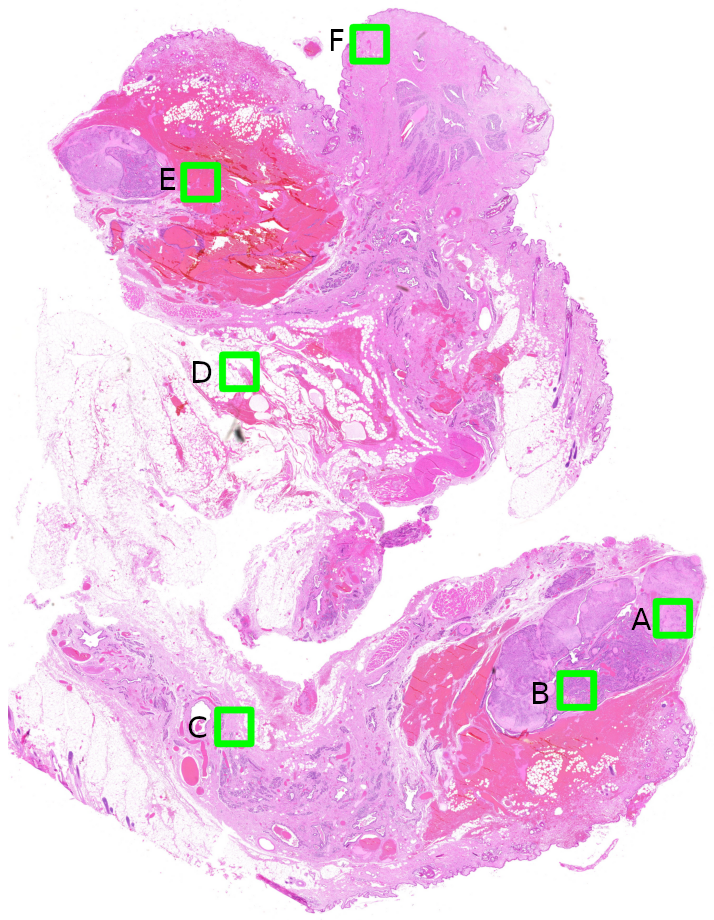}}
  \centerline{(a) HE slide}\medskip
\end{minipage}
\hfill
\begin{minipage}[b]{0.48\linewidth}
  \centering
  \centerline{\includegraphics[width=3.5cm]{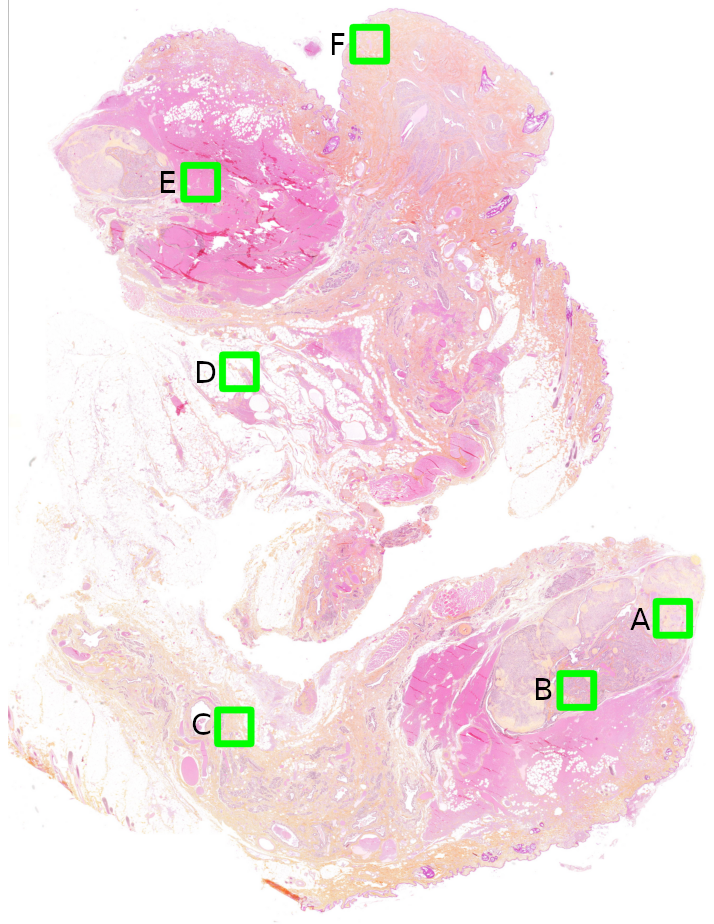}}
  \centerline{(b) HES slide}\medskip
\end{minipage}
\caption{\textbf{Slide Registration} - Example of a registered pair of HE-HES slides in the test set (first fold). The defined regions $A\text{-}F$ (2024x2024 pixels patches) are used in Figures 2 and 3.}
\end{figure}

\subsection{Quantitative analysis}
\label{ssec:unetperfs}

To evaluate our model on the regression task, we reported the Mean Absolute Error (MAE) on the 3-fold validation sets. We used three versions of this metric: $MAE$ computed over all the pixels, $MAE_{S}$ in the Saffron areas ($H_{S} > 0.05$) and $MAE_{B}$ in the other areas. We also used the $Dice$ score as semantic segmentation metric by getting masks by tresholding the Saffron concentrations. $mDice$ was computed as a mean of $Dice$ scores obtained using twenty thresholds from 0. to 1. and $Dice_{S}$ was obtained with a null threshold.

\begin{table}
\begin{center}
   \begin{tabular}{| r | r | r | r | r |}
   \hline $MAE$ & $MAE_{S}$ & $MAE_{B}$ & $mDice$ & $Dice_{S}$
   \tabularnewline
     \hline
     0.0668 & 0.1237 & 0.0303 & 0.6536 & 0.8097 \\
     $\pm$ .0002 &  $\pm$ .0003 & $\pm$ .0088 & $\pm$ .0003 & $\pm$ .0231 \\ \hline
   \end{tabular}
 \caption{\textbf{Pixel-wise regression evaluation}: Metrics evaluating the pixel-wise Saffron concentration prediction task obtained with a 3 folds-validation.}
 \end{center}
\end{table}

On the first validation fold, we get a lower $MAE_{S}$ when using the MSE Loss (0.1327) compared to the Weighted version (0.1240) showing that the wMSE loss allows the model to perform better on Saffron areas while having similar performances on the background ($MAE_{B}$: 0.013 (MSE) vs 0.022 (wMSE)).

As the quantification of collagen can have prognostic value in some pathologies, we tested the quantitative performances of our model on the different areas of the Figure 1 and some additional random test patches. For each region, we compared the global mean Saffron concentration between the prediction $mH_{\hat{S}}$ and the ground truth $m{H_{S}}$. Figure 2 shows that we have a strong linear relationship between our quantitative predictions and the ground truth, proving the effectiveness of our model to quantify collagen.

\subsection{Qualitative analysis}
\label{ssec:prediction}

In Figure 3, we show predictions on the four regions $A\text{-}D$ defined in Figure 1. The columns are in this order: the input HE patches $I_{HE}$, the estimated $H_{\hat{S}}$ and ground truths $H_{\hat{S}}$ Saffron concentrations $H_{\hat{S}}$, the reconstructed HES images $I_{HE+\hat{S}}$ and the registered HES patches $I_{HES}$. We can see that our model is able to closely predict the Saffron intensities $H_{\hat{S}}$ compared to the ground truths $H_{S}$ over very different areas. At the same time, this Figure shows that the reconstructed images $I_{HE+\hat{S}}$ successfully highlights the collagen while reasonably looking like a realistic HES image. The real HES images $I_{HES}$ seem to have weak Eosin concentrations - which might be caused by the restaining procedure - however this is not the case for $I_{HE+\hat{S}}$ which contains the H and E concentrations of the input HE image.

\begin{figure}
  \centering
  \centerline{\includegraphics[width=6cm]{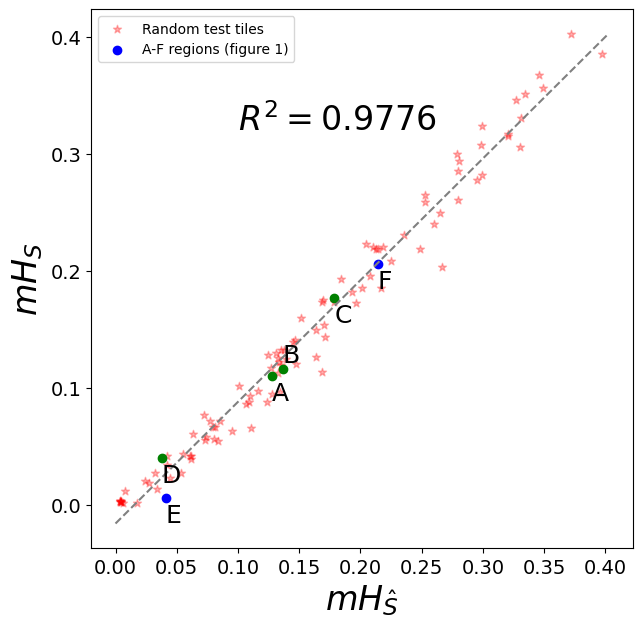}}
  \caption{\textbf{Quantitative Evaluation} - Least squares linear regression between the mean Saffron concentrations from the predictions $mH_{\hat{S}}$ and the ground truths $m{H_{S}}$ on test images (fold 1). Regions with green dots are the regions of Figure 3.}
\end{figure}


\begin{figure*}[t!]
  \hspace{1.3cm} $I_{HE}$\hfill$H_{\hat{S}}$\hfill$H_{S}$\hfill$I_{HE+\hat{S}}$\hfill$I_{HES}$\hspace{1.3cm} \\
  \includegraphics[width=3.45cm]{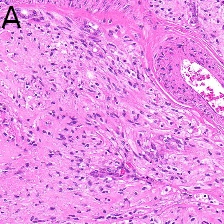}
  \hfill
  \includegraphics[width=3.45cm]{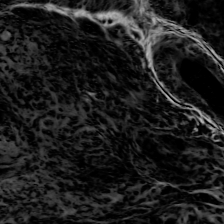}
  \hfill
  \includegraphics[width=3.45cm]{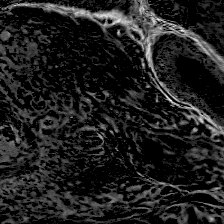}
  \hfill
  \includegraphics[width=3.45cm]{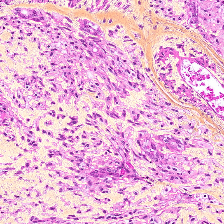}
  \hfill
  \includegraphics[width=3.45cm]{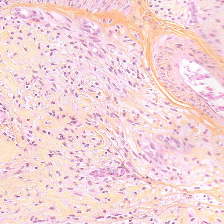}
  \includegraphics[width=3.45cm]{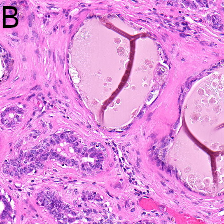}
  \hfill
  \includegraphics[width=3.45cm]{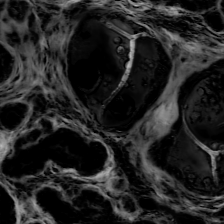}
  \hfill
  \includegraphics[width=3.45cm]{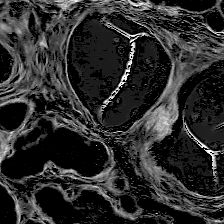}
  \hfill
  \includegraphics[width=3.45cm]{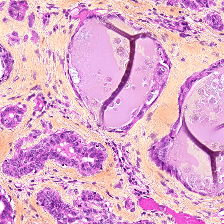}
  \hfill
  \includegraphics[width=3.45cm]{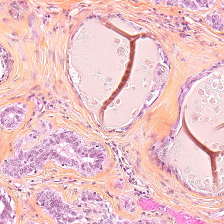}
  \includegraphics[width=3.45cm]{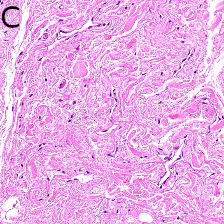}
  \hfill
  \includegraphics[width=3.45cm]{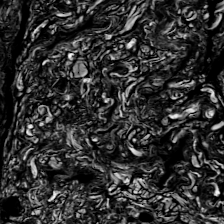}
  \hfill
  \includegraphics[width=3.45cm]{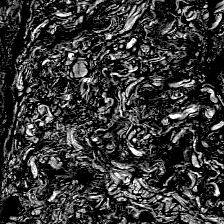}
  \hfill
  \includegraphics[width=3.45cm]{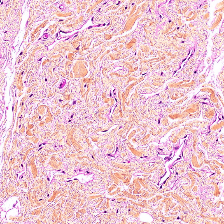}
  \hfill
  \includegraphics[width=3.45cm]{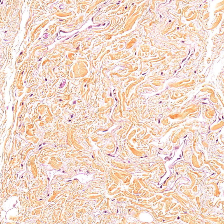}
  \includegraphics[width=3.45cm]{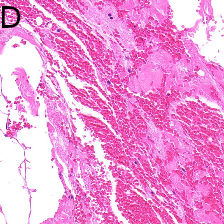}
  \hfill
  \includegraphics[width=3.45cm]{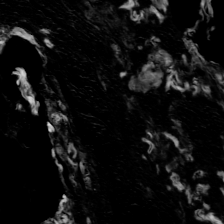}
  \hfill
  \includegraphics[width=3.45cm]{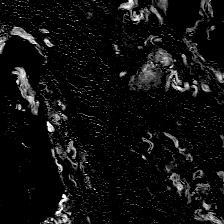}
  \hfill
  \includegraphics[width=3.45cm]{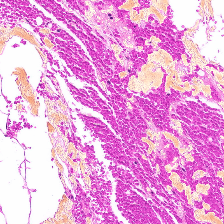}
  \hfill
  \includegraphics[width=3.45cm]{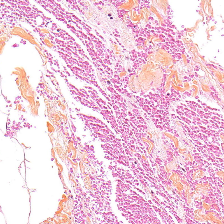}
  \caption{\textbf{Prediction Visualization} - The columns represent in this order: four input HE patches $I_{HE}$ extracted from the figure 1, the estimated Saffron concentrations $H_{\hat{S}}$, the ground truths Saffron concentrations $H_{S}$, the reconstructed HES images $I_{HE+\hat{S}}$ and the registered HES patches $I_{HES}$. Regions A and B are in tumor areas while C and D are benign regions.}
\end{figure*}
\vfill
\pagebreak

\section{Discussion \& Conclusion}
\label{sec:foot}

In this paper, we showed that it is possible to train a deep learning model that predicts Saffron staining from HE images, allowing to segment and to quantify the collagen which has both diagnostic and prognostic value for cancer malignancy.

We trained a UNet to predict Saffron concentrations - extracted via stain deconvolution - as a proxy task for predicting collagen, on a dataset of seven pairs of registered and restained HE-HES slides. The Table 1. reports the metrics of our model on a 3 fold validation sets, reaching a Mean Absolute Error of 0.0668 $\pm$ 0.0002 ($H_{S} \in [0, 1]$). The Figure 2 demonstrates the strong capability of our model to estimate the collagen quantity by showing the strong linear relationship ($R^2 = 0.9776$) of the mean Saffron concentrations between the predictions and the ground truths on test patches. Finally, in Figure 3 we show qualitative predictions, illustrating the convincing spatial Saffron concentration predictions along with reasonably realistic-looking HES reconstructions.

In future works it would be interesting to expand this methodology to a bigger multi-organs dataset, so we can train a general robust Saffron estimator model. We also have not discussed the domain shift between HE and HES. Our solution can be used as data augmentation and could improve on HES images the performances of the deep learning models that are most of the time trained on HE data. We hope this work can aid in improving the clinical workflow while reducing reagent costs for laboratories.
\newpage
\section{Compliance with Ethical Standards}
The dataset used in this work was created at the University of Veterinary Medicine Vienna with ethical permission.
\bibliographystyle{IEEEbib}
\bibliography{strings,refs}

\end{document}